\PassOptionsToPackage{table,xcdraw}{xcolor}
\documentclass[sigconf]{acmart}
\setcopyright{none}
\renewcommand\footnotetextcopyrightpermission[1]{} 

\acmYear{2025}

\usepackage{dingbat}
\usepackage{listings}
\usepackage{color}
\usepackage{xcolor}
\usepackage{multirow}
\usepackage{colortbl}
\usepackage{enumitem}

\usepackage[linesnumbered,ruled,boxed,lined]{algorithm2e}
\SetAlgoVlined 
\SetAlgoLined 
\SetKwInput{KwInput}{Input}  
\SetKwInput{KwOutput}{Output}  
\SetKwProg{Proc}{Procedure}{}{}

\DontPrintSemicolon

\usepackage{algpseudocode}
\usepackage{url}
\usepackage{float}
\usepackage{hyperref}
\usepackage{caption}
\usepackage{adjustbox}

\algtext*{EndProcedure}

\definecolor{mypurple}{HTML}{67379A}
\definecolor{myyellow}{HTML}{FAD978}
\definecolor{myblue}{HTML}{93AAD8}
\definecolor{myred}{HTML}{EB3323}
\definecolor{mygreen}{HTML}{4EAD5B}
\definecolor{greenbg}{rgb}{0.9, 1, 0.9}

\lstdefinestyle{mystyle}{
    backgroundcolor=\color{white},   
    basicstyle=\ttfamily\footnotesize,
    breakatwhitespace=false,         
    breaklines=true,                 
    captionpos=b,                    
    keepspaces=true,                 
    numbers=left,                    
    numbersep=5pt,                  
    showspaces=false,                
    showstringspaces=false,
    showtabs=false,                  
    tabsize=2,
    numberstyle=\tiny\color{black},
    framexleftmargin=5mm, frame=single, framerule=0pt,
    xleftmargin=5mm,
    xrightmargin=3mm,
    framesep=5mm, 
    rulecolor=\color{black},
    postbreak=\mbox{\textcolor{red}{$\hookrightarrow$}\space},
}
\usepackage{amsthm}
\theoremstyle{definition}
 
\theoremstyle{plain}

\newcommand{\lgtm}{\textsc{VulZoo}\xspace}

\usepackage{pifont}

\makeatletter
\newcommand{\removelatexerror}{\let\@latex@error\@gobble}
\makeatother

\begin{document}

\title{\lgtm: A Comprehensive Vulnerability Intelligence Dataset}

\author{Bonan Ruan$^\dagger$\hspace{0.2em} Jiahao Liu$^\dagger$\hspace{0.2em} Weibo Zhao$^\dagger$\hspace{0.2em} Zhenkai Liang$^\dagger$}
\affiliation{%
  \institution{\vspace{0.1cm}$^\dagger$\ National University of Singapore}
  \city{}
  \country{}
}
\vspace{0.1cm}
\email{{r-bonan, jiahao99, weibo, liangzk}@comp.nus.edu.sg}

\renewcommand{\shortauthors}{Bonan Ruan, Jiahao Liu, Weibo Zhao, and Zhenkai Liang}

\begin{abstract}

Software vulnerabilities pose critical security and risk concerns for many software systems.
Many techniques have been proposed to effectively assess and prioritize these vulnerabilities before they cause serious consequences.
To evaluate their performance, these solutions often craft their own experimental datasets from limited information sources, such as MITRE CVE and NVD, lacking a global overview of broad vulnerability intelligence.
The repetitive data preparation process further complicates the verification and comparison of new solutions.
To resolve this issue, in this paper, we propose \lgtm, a comprehensive vulnerability intelligence dataset that covers 17 popular vulnerability information sources.
We also construct connections among these sources, enabling more straightforward configuration and adaptation for different vulnerability assessment tasks (e.g., vulnerability type prediction).
Additionally, \lgtm provides utility scripts for automatic data synchronization and cleaning, relationship mining, and statistics generation.
We make \lgtm publicly available and maintain it with incremental updates to facilitate future research.
We believe that \lgtm serves as a valuable input to vulnerability assessment and prioritization studies.
The dataset with utility scripts is available at \href{https://github.com/NUS-Curiosity/VulZoo}{https://github.com/NUS-Curiosity/VulZoo}.
\end{abstract}



\maketitle

\section{Introduction}
\label{sec:intro}
Software vulnerabilities (SVs) have been growing in both scale and complexity, posing substantial security and economic risks in recent years.
In 2023 alone, the National Vulnerability Database (NVD)~\cite{nvd} cataloged 28,831 vulnerabilities, marking a 15\% increase from the previous year~\cite{vulr}.
Cybersecurity Ventures also estimated that cyberattacks cost the global economy approximately 8 trillion USD in the same year~\cite{cost}.
Swift response and remediation of reported SVs before exploitation are crucial for mitigating potential losses.
However, due to the large volume of vulnerabilities, it is infeasible to address all of them simultaneously.
Therefore, it is essential to accurately assess and prioritize exposed vulnerabilities, enabling developers and researchers to address the most critical SVs as promptly as possible.

Existing vulnerability assessment and prioritization techniques can be classified into two categories:
(1) \textit{Improving SV information quality}.
A software vulnerability can be profiled from multi-channel information, such as government vulnerability repositories, vulnerability-disclosure email lists, and forums.
These solutions aim to enhance the quality of information to better profile SVs.
For example, they address issues like inconsistency~\cite{cvssInconsistencies}, incorrectness~\cite{ruan2024kernjc}, and incompleteness~\cite{guo2022detecting}, as well as align different sources of information~\cite{qin2023vulnerability}.
(2) \textit{Exploring SV intrinsic property}.
These methods aim to examine SVs from multiple perspectives, such as assessing the  exploitability~\cite{you2017semfuzz}, predicting their vulnerable type~\cite{pan2023fine}, their impact~\cite{elbaz2020fighting}, and their inherent weaknesses~\cite{wu2020precisely}.

To evaluate these proposed approaches, researchers often craft specific datasets to check whether they achieve the design objectives.
However, there are two shortcomings of the current data preparation process.
Firstly, this process is sophisticated and time-consuming, which impedes the rapid verification and comparison of new approaches or tools for vulnerability assessment and prioritization.
Secondly, the data sources adopted in existing studies are mainly Common Vulnerabilities and Exposures (CVE) databases and software projects, which are limited compared to the large-scale open vulnerability intelligence available on the Internet.
Intuitively, a well-structured and well-maintained dataset can better profile vulnerabilities and aid in assessing and prioritizing them.
Unfortunately, to the best of our knowledge, no such dataset currently satisfies these requirements.

To bridge this gap, we introduce \lgtm, a comprehensive vulnerability intelligence dataset, which is readily usable and adaptable for broad vulnerability investigation tasks.
\lgtm is derived from 17 heterogeneous data sources, including structural and non-structural data. 
The structural data contains 604,943 \textbf{CVE records} from 4 popular online databases (MITRE CVE~\cite{cveWebsite}, NVD~\cite{nvd}, ZDI Advisory~\cite{zerodayinitiativeHomeZero}, and GitHub Advisory~\cite{githubGitHubAdvisory}), \textbf{assessment} related metrics from 8 vulnerability related catalogs (CPE~\cite{cpe}, CWE~\cite{mitreCommonWeakness}, CVSS~\cite{cvss}, KEV~\cite{cisaKnownExploited}, CAPEC~\cite{mitreCAPECCommon}, AttackerKB~\cite{attackerkb}, ATT\&CK~\cite{mitreMITREATTampCKxAE}, and D3FEND~\cite{mitreMITRED3FEND}).
The non-structural data involves the text of 46,882 vulnerability-related \textbf{mails}, 46,540 Proof of Concepts (\textbf{PoCs}), and 12,626 \textbf{patches}.
In addition, we mine 11 different \textbf{relationships} describing the connectives across data in different categories.
Detailed information is provided in \autoref{tab:data_statistics} and \autoref{tab:rel_statistics}.

In summary, we make the following contributions:

\begin{itemize}[leftmargin=10pt]
    \item Comprehensive Dataset. We build the \lgtm dataset (\textasciitilde6 GB) based on 17 heterogeneous vulnerability information sources with a well-designed structure.
    \item Utility scripts. We develop a set of scripts for data synchronization and cleaning, relationship mining, and statistics generation. These scripts are leveraged to keep \lgtm up-to-date.
\end{itemize}

\section{Dataset Construction}
\label{sec:construction}

To ensure that \lgtm can comprehensively describe the broad intelligence of software vulnerabilities and support various related tasks, we perform a thorough review of research and industry communities to include as many information channels as possible for profiling vulnerabilities.
In the subsequent sections, we first present the data sources included in \lgtm and then detail the data collection process.

\begin{table}[t]
  \centering
  \setlength{\belowcaptionskip}{+10pt}
  \caption{Data sources considered when constructing \lgtm}
  \begin{adjustbox}{width=0.9\linewidth, center}
  \begin{tabular}{l|l}
    \hline
    \textbf{Source}  & \textbf{Data of Interest} \\
    \hline
    MITRE CVE & CVE records \\
    NVD & Enhanced CVE records and CPE dictionary \\
    ZDI Advisory & Enhanced CVE records\\
    GitHub Advisory & Enhanced CVE records \\
    CISA KEV & Records of known exploited CVEs \\
    MITRE CWE & Weakness catalog \\
    MITRE CAPEC & Attack pattern catalog \\
    MITRE ATT\&CK & Offensive technique catalog \\
    MITRE D3FEND & Defensive technique catalog \\
    Rapid7 AttackerKB & Crowdsourcing CVE assessments \\
    OffSec Exploit-DB & Proof of Concepts (PoCs) \\
    Bugtraq & Mails on CVEs \\
    Full-Disclosure & Mails on CVEs \\
    OSS-Security & Mails on CVEs in open-source projects \\
    Linux-CVE-Announce & Mails on CVEs in Linux kernel \\
    GitHub & Patches of CVEs \\
    git.kernel.org & Patches of CVEs in Linux kernel \\
    \hline
  \end{tabular}
  \end{adjustbox}
  \label{tab:data_sources}
  \vspace{-0.5cm}
\end{table}

\subsection{Data Sources}
\label{sec:data_sources}

\lgtm adopts 17 data sources for the dataset construction, their names along with the available data of interest presented in \autoref{tab:data_sources}. We give a brief introduction to each data source as follows:

\begin{itemize}[leftmargin=10pt]
    \item MITRE CVE~\cite{cveWebsite} is the official CVE database for cataloging publicly disclosed vulnerabilities in cybersecurity. Each entry in this database includes a unique identifier for a specific vulnerability, along with textual descriptions and corresponding links.
    \item NVD is another critical vulnerability database maintained by the U.S. government. NVD performs analysis for published CVEs in MITRE CVE database and provides more detailed vulnerability intelligence, such as impact metrics (Common Vulnerability Scoring System, CVSS~\cite{cvss}), vulnerability types (Common Weakness Enumeration, CWE~\cite{mitreCommonWeakness}), applicability statements (Common Platform Enumeration, CPE~\cite{cpe}), and other pertinent metadata. NVD also maintains the CPE dictionary for reference. 
    \item ZDI Advisory~\cite{zerodayinitiativeHomeZero} maintains a list of disclosed vulnerabilities acquired by Zero Day Initiative~\cite{zerodayinitiativeHomeZero} (one of the world's largest vendor-agnostic bug bounty programs), which provides CVE information and CVSS assessment.
    \item GitHub Advisory~\cite{githubGitHubAdvisory} is a new vulnerability database inclusive of CVEs and GitHub-originated security advisories beginning in 2017. This database provides CVE information, CVSS and CWE assessment, and affected packages with version ranges specified.
    \item CISA KEV~\cite{cisaKnownExploited} is a catalog of Known Exploited Vulnerabilities maintained by CISA. CVEs in this list are authoritatively confirmed to be exploited in the wild. Hence, KEV serves as a valuable input to vulnerability prioritization tasks.
    \item MITRE CWE~\cite{mitreCommonWeakness} is a hierarchically organized list specialized for the root cause mapping of vulnerabilities. Records in vulnerability databases (\textit{e.g.,} NVD and GitHub Advisory) are correlated CWE entries via CWE identifiers.
    \item MITRE CAPEC (Common Attack Pattern Enumeration and Classification)~\cite{mitreCAPECCommon} provides a dictionary of known attack patterns employed by adversaries to exploit known cyber weaknesses.
    \item MITRE ATT\&CK (Adversarial Tactics, Techniques, and Common Knowledge)~\cite{mitreMITREATTampCKxAE} is a knowledge base of adversary tactics and techniques based on real-world observations. ATT\&CK has been widely used in threat modeling~\cite{huang2021open,ahmed2022mitre}. Many techniques in ATT\&CK have relationships with CAPEC entries.
    \item MITRE D3FEND~\cite{mitreMITRED3FEND} serves as the defensive counterpart framework against ATT\&CK, providing a countermeasure knowledge base derived from research and development literature.
    \item Rapid7 AttackerKB~\cite{attackerkb} is a crowdsourcing knowledge base, where security researchers contribute detailed assessment analysis or attach tags (such as \textit{Easy to weaponize}, \textit{Gives privileged access}, and \textit{Vulnerable in default configuration}) to vulnerabilities.
    \item OffSec Exploit-DB~\cite{exploitdb} is a CVE-compliant repository for exploits and proof of concepts\footnote{In this paper, we do not distinguish between exploits and proof of concepts, but refer to them as PoCs.}. 
    \item Bugtraq~\cite{openwallBugtraqMailing} was a mailing list dedicated to cybersecurity issues, including vulnerabilities, vendor announcements, exploitations, and patches. It was shut down on January 31, 2021.
    \item Full-Disclosure~\cite{openwallFulldisclosureMailing} is another cybersecurity-related mailing list for discussions and vulnerability disclosures. It got a temporary interruption but resumed in 2014.
    \item OSS-Security~\cite{openwallOsssecurityMailing} is a mailing list focusing on security issues in open-source projects.
    \item Linux-CVE-Announce~\cite{kernelLinuxcveannouncevgerkernelorgArchive} is a mailing list specialized for announcements of CVEs in Linux kernel.
    \item GitHub~\cite{githubGitHubLets} hosts numerous open-source projects, as well as the patches for vulnerabilities discovered in these projects.
    \item git.kernel.org~\cite{kernelKernelorgRepositories} hosts the source code repositories (including common commits and patches) of the Linux kernel.
\end{itemize}

\begin{figure}[t]
  \centering
  \includegraphics[width=0.7\linewidth]{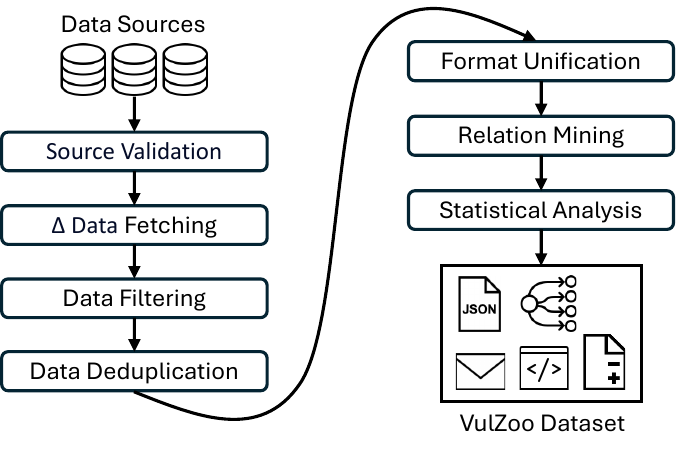}
  \vspace{-0.2cm}
  \caption{Dataset Construction Process}
  \label{fig:workflow}
  \vspace{-0.3cm}
\end{figure}

\subsection{Data Collection}
\label{sec:data_collection}

Figure~\ref{fig:workflow} illustrates the high-level overview of \lgtm's data collection process, demonstrating how raw data is transformed into a well-structured dataset.
We introduce the process step by step.

\noindent\textbf{Source validation and data fetching.} 
The data sources for \lgtm are diverse and come in different formats.
To handle them, we have customized four crawling strategies.
(1) For sources with \textit{Git-based repositories}, such as MITRE CVE, NVD, ZDI Advisory, GitHub Advisory, MITRE ATT\&CK, OffSec Exploit-DB, and Linux-CVE-Announce, we configure them as \textit{git submodules} within \lgtm repository, benefiting from the natural incremental data synchronization feature of Git.
(2) For sources releasing files via \textit{stable hyperlinks}, \textit{i.e.,} NVD (for CPE dictionary), CISA KEV, CWE, CAPEC, and D3FEND, we directly download the target files and decompress them if necessary. 
(3) For sources containing \textit{discrete data with respective links}, \textit{i.e.,} Bugtraq, Full-Disclosure, and OSS-Security, we develop a mail crawler to first retrieve and parse mail links for each mail, and then download them individually. 
(4) For patches on \textit{GitHub and git.kernel.org}, we develop a program to identify patch links in NVD by tags and then fetch the contents. 

During the preliminary analysis, we observed many links in mailing lists and NVD are invalid. 
Additionally, there are still 1,605 links tagged as \textit{Patch} in NVD pointing to www.securityfocus.com, which has been inactive for a long time. 
Therefore, our crawlers first validate the link status before proceeding with further operations.

\noindent \textbf{Data filtering, deduplication, and format unification.}
After fetching the raw data, we filter out irrelevant files (\textit{e.g.,} READMEs and CHANGELOGs) and retain only the vulnerability-related data to improve dataset readiness for programming usage. 
Following data filtering, we conduct SHA256-based file comparison for deduplication. 
We observe that there are respectively 71, 24, and 6 duplication cases among mails from Bugtraq, Full-Disclosure, and OSS-Security mailing lists. 
After deduplication, we convert all structural data from various formats into JSON to provide a unified data interface for future usage.

\noindent \textbf{Relationship mining and statistical analysis.}
After the aforementioned processing, we get data in 4 types: structural JSON files, non-structural mails, patches, and PoC code snippets.
To profile vulnerabilities from different aspects, we mine the relationships within the data to unveil and enhance data interconnection.
Note that we deal with real-world vulnerabilities, not synthetic ones. 
As a result, not all CVEs have the same number of relationships to the other data.
For example, PoCs do not exist for the majority of existing CVEs. 
We mine 11 types of relationships in all, covering all the edges in the topology graph shown in \autoref{fig:topology}.
Once the dataset is ready, we generate the statistics for inspection.

\section{Dataset Description}
\label{sec:description}

In this section, we first provide a global overview of \lgtm's contents.
Following that, we take CVE-2020-7247 as an example to demonstrate how the assembled intelligence in \lgtm enables comprehensive profiling for vulnerabilities.

\subsection{Dataset Overview}
\label{dataset_overview}

\begin{table}[t]
  \centering
  \setlength{\belowcaptionskip}{+10pt}
  \caption{Descriptive Statistics of Data in \lgtm}
  \begin{adjustbox}{width=0.96\linewidth, center}
  \begin{tabular}{l|l|l|l}
    \hline
\textbf{Format}                  & \textbf{Category}                                    & \textbf{Measurement}                                         & \textbf{Value} \\ \hline
                                 &                            & \cellcolor[HTML]{F89588}MITRE-recorded CVEs                  & 320,861        \\
                                 &                             & \cellcolor[HTML]{F89588}NVD-recorded CVEs                    & 253,722        \\
                                 &                             & \cellcolor[HTML]{F89588}ZDI-recorded CVEs                    & 13,291         \\
                                 & \multirow{-4}{*}{CVE Record} & \cellcolor[HTML]{F89588}GitHub-recorded CVEs                 & 17,069         \\ \cline{2-4} 
                                 &                                                      & \cellcolor[HTML]{F0F8FF}CPE Names                            & 1,271,275      \\
                                 &                                                      & \cellcolor[HTML]{E6F7FF}CWE Weaknesses                       & 963            \\
                                 &                                                      & \cellcolor[HTML]{CCEFFF}CVSS Metrics                         & N/A            \\
                                 &                                                      & \cellcolor[HTML]{B3E6FF}KEV                                  & 1,120          \\
                                 &                                                      & \cellcolor[HTML]{99DDFF}AttackerKB Assessments               & 1,665          \\
                                 &                                                      & \cellcolor[HTML]{80D4FF}CAPEC Attack Patterns                & 615            \\
                                 &                                                      & \cellcolor[HTML]{66CCFF}ATT\&CK Techniques                   & 1062           \\
\multirow{-12}{*}{Structural}    & \multirow{-8}{*}{Assessment}                         & \cellcolor[HTML]{4DC3FF}D3FEND Techniques                    & 183            \\ \hline
                                 & PoC                                                  & \cellcolor[HTML]{CBCEFB}Exploit-DB PoCs                      & 46,540         \\ \cline{2-4} 
                                 &                                                      & \cellcolor[HTML]{F8CB7F}CVE-related Bugtraq Mails            & 17,404         \\
                                 &                                                      & \cellcolor[HTML]{F8CB7F}CVE-related Full-Disclosure Mails    & 12,448         \\
                                 &                                                      & \cellcolor[HTML]{F8CB7F}CVE-related OSS-Security Mails       & 14,976         \\
                                 & \multirow{-4}{*}{Mail}                               & \cellcolor[HTML]{F8CB7F}CVE-related Linux-CVE-Announce Mails & 2,054          \\ \cline{2-4} 
\multirow{-6}{*}{Non-structural} & Patch                                                & \cellcolor[HTML]{76DA91}Patch Files                          & 12,626         \\ \hline
\end{tabular}%
  \end{adjustbox}
  
  \label{tab:data_statistics}
  \vspace{-0.5cm}
\end{table}

\lgtm organizes the structural and non-structural data into 5 categories: \textit{CVE Record}, \textit{Assessment}, \textit{PoC}, \textit{Mail}, and \textit{Patch}. 
The statistics for each category as of June 13, 2024, are presented in Table~\ref{tab:data_statistics}. 
We give a brief introduction to each category as follows:


\begin{itemize}[leftmargin=10pt]
\item CVE Record. Each CVE record contains basic information regarding a CVE ID, such as descriptions, names, and versions of affected products.  
\lgtm integrates 4 different CVE databases as data sources.
Thus, there could be one or more pieces of information describing the same CVE.
%
This benefits future research from two aspects.
First, CVE instances from different databases contribute more or less unique information, such as CPE metrics in NVD, and independent CVSS assessment in GitHub Advisory. 
The composition works better than any single database. 
Second, since the information describing a CVE comes from various sources, \lgtm inherently provides a valuable resource for CVE differential analysis and alignment research.

\item Assessment. This category includes 8 different vulnerability related assessment metrics. 
Specially, CPE is a huge structured naming scheme, commonly used to delineate the vulnerable products and version ranges regarding a CVE ID. 
Thanks to the rapid development of modern software engineering, CPE dictionary contains more than one million today and still grows very fast. 
CVSS is a special assessment metric used to capture the principal characteristics of a vulnerability as a vector string and produce a numerical score to reflect its severity. The vector string and the score are usually attached to CVE records.
Hence, there is no value for CVSS in \autoref{tab:data_statistics}. 
%
%
For clarity, we omit the introductions to CWE, KEV, CAPEC, AttackerKB, ATT\&CK, and D3FEND, as they have been introduced in \autoref{sec:data_sources}. 

\item PoC. Most of the PoC files are code snippets that could be executed to trigger a specific vulnerability. Some others may describe the manual process of reproducing a vulnerability.

\item Mail. Each mail is stored as a text file recording the communications related to one or more CVEs.

\item Patch. Each patch file is the code difference generated by one commit operation, with optional description messages.
\end{itemize}

\begin{figure}[t]
  \centering
  \includegraphics[width=0.9\linewidth]{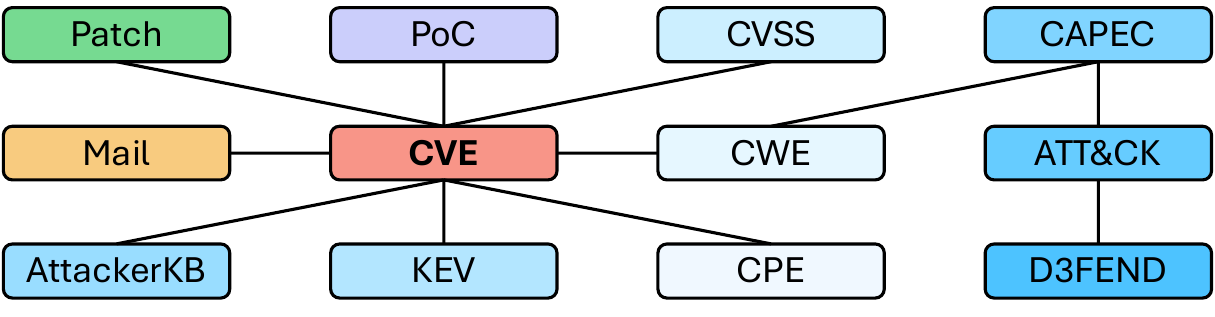}
  \caption{Topology of \lgtm Dataset}
  \label{fig:topology}
\end{figure}

\begin{table}[t]
  \centering
  \setlength{\belowcaptionskip}{+10pt}
  \caption{Descriptive Statistics of Relationships in \lgtm}
  \begin{adjustbox}{width=0.7\linewidth, center}
  \begin{tabular}{l|l}
    \hline
\textbf{Relationship}            & \textbf{Number} \\ \hline
CVEs with CPE Names              & 224,998         \\
CVEs with CWE Weaknesses         & 179,950         \\
CVEs with CVSS Metrics           & 234,260         \\
CVEs mentioned in KEV            &   1,120              \\
CVEs with AttackerKB Assessments & 1,108           \\
CVEs with Exploit-DB PoCs        & 24,587          \\
CVEs with Mails                  & 42,030          \\
CVEs with Patch Files            & 10,548          \\
CWE -- CAPEC                     & 336 / 450       \\
CAPEC -- ATT\&CK                 & 177 / 36        \\
ATT\&CK -- D3FEND                & 301 / 121       \\ \hline
\end{tabular}%
  \end{adjustbox}
  
  \label{tab:rel_statistics}
  \vspace{-0.3cm}
\end{table}

Table~\ref{tab:rel_statistics} presents the 11 relationships mined from \lgtm.
Specifically, the first 8 relationships center on CVE records and the last 3 ones are bipartite relationships among CWE, CAPEC, ATT\&CK, and D3FEND.
For example, there are 336 entries in the CWE catalog referring to CAPEC and 450 entries in the CAPEC catalog referring to CWE.
For a better understanding of the relationship of data in \lgtm, we draw their connectivity as a topology graph, as shown in Figure~\ref{fig:topology}.
Here, nodes are the data listed in Table~\ref{tab:data_statistics} and edges denote the relationship in Table~\ref{tab:rel_statistics}.

\begin{figure}[t]
  \centering
  \includegraphics[width=1\linewidth]{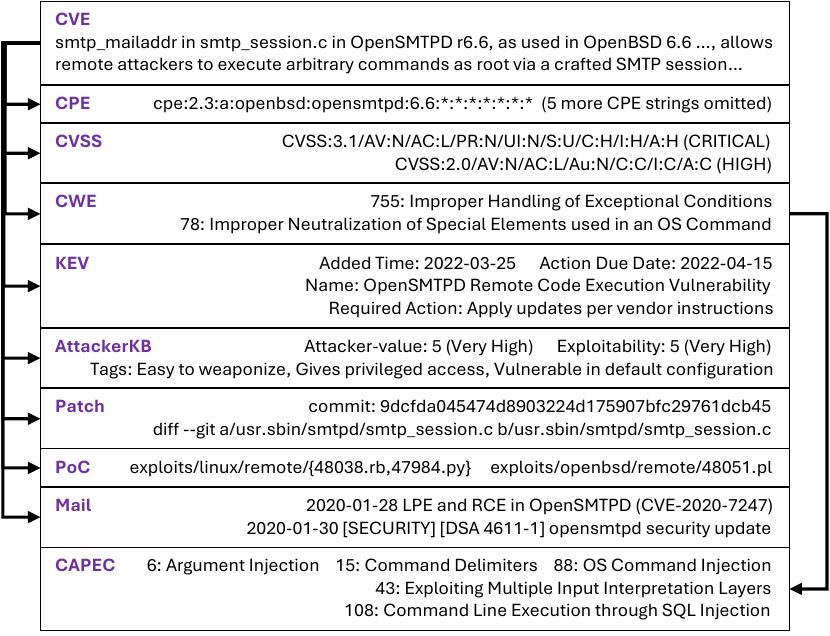}
  \caption{CVE-2020-7247 Intelligence in \lgtm}
  \label{fig:example}
  \vspace{-0.2cm}
\end{figure}

\subsection{Example: CVE-2020-7247}

We employ CVE-2020-7247 to demonstrate \lgtm's vulnerability profiling capability.
Starting from the \textit{CVE} node in~\autoref{fig:topology}, we traverse the topology graph to collect all information in \lgtm describing it.
\autoref{fig:example} shows the summarized result for this investigation, covering all nodes in \autoref{fig:topology} except for ATT\&CK and D3FEND. 
From the assembled intelligence, we know that CVE-2020-7247 is a critical command injection vulnerability in OpenSMTPD, with one patch file (modifying \verb|smtp_session.c|) and three PoCs (written in Ruby, Python, and Perl) available. 
Researchers mark this CVE as having very high value and exploitability because it exists in the default configuration, is easy to exploit, and grants privileged access once exploited.
CISA added this CVE into KEV and required remediation. Two mails contain the vulnerability analysis and remediation details. We argue that such an information composition profiles CVE-2020-7247 better than any single database.

\section{Application Scenarios}
\label{sec:application}

Thanks to \lgtm's large-scale data and heterogeneous characteristics, this dataset can support a wide range of vulnerability assessment and prioritization research.
In this section, we present three application scenarios of \lgtm.

\noindent \textbf{Severity and Type Assessment.} Researchers can train statistical or deep learning models based on the extensive data in \lgtm to predict the CVSS score and vector for a new vulnerability. 
This goal can also be achieved by proposing and applying similarity (distance) based algorithms to measure the similarity (distance) between vulnerabilities in \lgtm and future ones. 
These approaches also can be applied to predict CWEs of unseen vulnerabilities.

\noindent \textbf{Intelligence Alignment.} As \lgtm has already integrated vulnerability information from multiple sources, researchers can directly conduct alignment studies on this dataset.
For example, they can mine deep relationships and resolve inconsistencies among both structural and non-structural entities in \lgtm.

\noindent \textbf{Information Augmentation.} To facilitate the vulnerability management process, researchers can augment the vulnerability descriptions by deriving key aspects, such as affected files and exploitation status, from \lgtm, if they are missing by human analysts.

\section{Conclusion}
\label{sec:conclusion}

This paper presents \lgtm, a comprehensive vulnerability intelligence dataset containing heterogeneous vulnerability information derived from 17 data sources.
For easier updates and usage, we have developed multiple utility scripts to automate data synchronization and cleaning, relationship mining, and statistics generation processes. 
We believe that future vulnerability assessment and prioritization research will benefit from our dataset.


\bibliographystyle{ACM-Reference-Format}
\bibliography{ref}


\end{document}